\documentclass[prl,aps,twocolumn,floatfix]{revtex4}
\usepackage{amsmath,graphics,epsfig,color,verbatim}
\begin{document}

\title{Electronic coherence in $\delta$-Pu: A DMFT study}

\author{C. A. Marianetti$^1$, K. Haule$^2$, G. Kotliar$^2$, and M.J. Fluss$^1$ }

\date{\today}

\begin{abstract}
A combination of Density Functional Theory and the Dynamical Mean Field theory
(DMFT) is used to calculate the magnetic susceptibility, heat capacity, and the
temperature dependence of the valence band photoemission spectra.  The
continuous-time hybridization expansion quantum Monte-Carlo is utilized to
provide the first approximation-free DMFT solution of \emph{fcc} $\delta$-Pu
which includes the full rotationally-invariant exchange interaction.  We
predict that $\delta$-Pu has a Pauli-like magnetic susceptibility near ambient
temperature, as in experiment, indicating that electronic coherence causes the
absence of local moments.  Additionally, we show that volume expansion causes a
crossover from incoherent to coherent electronic behavior at increasingly lower
temperatures.  
  
\end{abstract}

\address{$^1$ Lawrence Livermore National Laboratory, Livermore, CA}
\address{$^2$ Department of Physics and Astronomy and Center for Condensed Matter
Theory, Rutgers University, Piscataway, NJ 08854--8019}

\maketitle

Pu embodies a forefront of both technology and theoretical condensed-matter
physics.  Elemental Pu displays exotic physical behavior that continues to
defy explanation. For example, it exhibits six allotropic phases at ambient
pressure, the low-density \emph{fcc} $\delta$ phase has a negative coefficient
of thermal expansion, and the volume expands by more than $25\%$ when the
system is heated from the high-density monoclinic $\alpha$ phase to the
$\delta$ phase. Even many of the basic electronic properties have not been
definitively characterized and understood.  Regarding the pure $\delta$ phase,
a complicating factor is that it is only stable in the temperature range
$580K<T<700K$.  However, the $\delta$ phase can be stabilized at low
temperatures by a variety of alloying elements such as Ga and Am.  This allows
for the experimental exploration of the $\delta$ phase at low temperatures,
with the caveat that it is not clear what changes the alloying element may be
inducing.  

Lashley \emph{et al.} \cite{Lashley:2005} have measured the magnetic
susceptibility to be Pauli-like in both the $\alpha$ and $\delta$
phases, and hence detect no presence of localized magnetic moments.
Similarly, Heffner \emph{et
  al.}\cite{Heffner:2006,Heffner:2006B,Heffner:2007} have used $\mu
SR$ and showed that there is no ordered magnetic moments in $\alpha$
Pu nor in $\delta$-stabilized Pu (i.e. 4.3\% Ga) for temperatures down
to $4\,$K.  Nuclear magnetic resonance (NMR) measurements by Curro
\emph{et al.}\cite{Curro:2004} also show an absence of magnetic
moments.

The linear coefficient of the specific heat for $\delta$-stabilized Pu has been
measured by various groups and the resulting values are $42\frac{mJ}{mol\,
K^2}$ for alloying with 2\% of Ga\cite{Lashley:2005}, $64\frac{mJ}{mol\,K^2}$
for 5\% of Al\cite{Lashley:2003}, and $35-55\frac{mJ}{mol\,K^2}$ for 8-20\% of
Am\cite{Javorsky:2006}.  The large variation among these measurements may be
due to the fact that the $\delta$ phase has been stabilized by a different
alloying element in each study. Furthermore, Pu easily incorporates elements
such as Fe during synthesis, while radiation self-damage may generate further
defects that may influence the low-temperature specific heat.

Several previous studies have applied a combination of Density Functional
Theory and the Dynamical Mean Field Theory (DFT+DMFT) \cite{Kotliar:2006} to
$\delta$ Pu.  DMFT requires a solution of an auxiliary quantum impurity
problem, and for the corresponding impurity model of Pu, no exact method was
available in the past.  Savrasov \emph{et al.}\cite{Savrasov:2001} used an
interpolative solver to calculate the energy and the photoemission spectra of
Pu.  The approach yielded a significant improvement for the volume of the
$\delta$ phase of Pu compared to DFT.  Pourovskii \emph{et al.}
\cite{Pourovskii:2007} computed the photoemission spectra and the heat capacity
using the Fluctuation Exchange Approximation (FLEX) as an impurity solver.  The
applicability of FLEX to Pu is questionable given the strongly correlated
nature of $\delta$-Pu.  Zhu \emph{et al.}\cite{Zhu:2007} computed the
photoemission spectra using the the Hirsch-Fye quantum Monte-Carlo impurity
solver and show that the occupation of the $f$ orbital is close to $n_f\sim 5$.
However, the limitation of this method to treat the realistic atomic multiplet
structure (Hund's rule coupling)  \cite{Kotliar:2006} precludes a precise
description of the problem.  Shim \emph{et al.} \cite{Shim:2007B} predicted the
photoemission spectra, the X-ray absorption spectroscopy branching ratio, and
the mixed-valence nature of Pu.  Although the mixed-valence state was
identified in ref.~\onlinecite{Shim:2007B}, the temperature and pressure
dependence of the electronic state was not addressed.

In this letter, we demonstrate the absence of magnetic moments in $\delta$-Pu
by computing the magnetic susceptibility as a function of temperature.  We show
that expanding the Pu lattice results in an incoherent metallic state with 
Curie-Weiss susceptibility at increasingly lower temperatures.  Additionally,
we elucidate the nature of the the mixed-valence state by predicting the
temperature dependence of the photoemission spectra in an extended temperature
regime starting below ambient  and continuing beyond the melting temperature.

DMFT maps the interacting lattice problem onto an impurity problem where the
non-interacting bath function is determined self-consistently
\cite{Georges:1996}.  The effective impurity problem is then solved using the
continuous-time quantum Monte-Carlo (CTQMC) method
\cite{Werner:2006,Haule:2007}.  More specifically, the recently developed
hybridization expansion CTQMC method is used to exactly sum the diagrams
resulting from expansion in powers of the hybridization strength between the
Pu atom and the DMFT fermionic bath \cite{Haule:2007}.  This method allows one
to include the full rotationally-invariant exchange interaction without
approximation, in contrast to Hirsch-Fye QMC\cite{Kotliar:2006}.  Although our
implementation of CTQMC is extremely efficient, massive parallel computer
resources are required to solve the 14 orbital DMFT impurity problem for Pu.
The Atlas supercomputer at Lawrence Livermore National Lab was used to perform
the calculations, and time was awarded under the Atlas grand challenge
program.

DFT+DMFT calculations were performed using an orthogonalized LMTO basis which
provides $f$ orbitals that have a maximal $f$ character\cite{Toropova:2007}.
In accordance with previous studies, the on-site Coulomb repulsion was chosen
to be $U=4eV$ and, the Slater integrals ($F_2$, $F_4$, and $F_6$) were computed
using an atomic physics code \cite{Cowan:1981} and rescaled by 80\% to account
for screening in the solid. Summations over the first Brillioun zone were
performed with 15x15x15 meshes, with the exception of the heat capacity
calculations where up to 40x40x40 meshes were used to ensure convergence.  Pu
valencies of $N_f=4-7$ were retained in the QMC simulation, and retaining
higher/lower valencies had no appreciable influence on the results. The total occupation of
the Pu-$5f$ orbitals in this study is approximately $n_f=5.2$, consistent with
previous calculations\cite{Shim:2007B} and electron energy loss
experiments of the \emph{N} edge\cite{Moore:2007} and the \emph{O} edge\cite{Butterfield:2008}.

\begin{figure}[htb]
\includegraphics[width=\linewidth,clip= ]{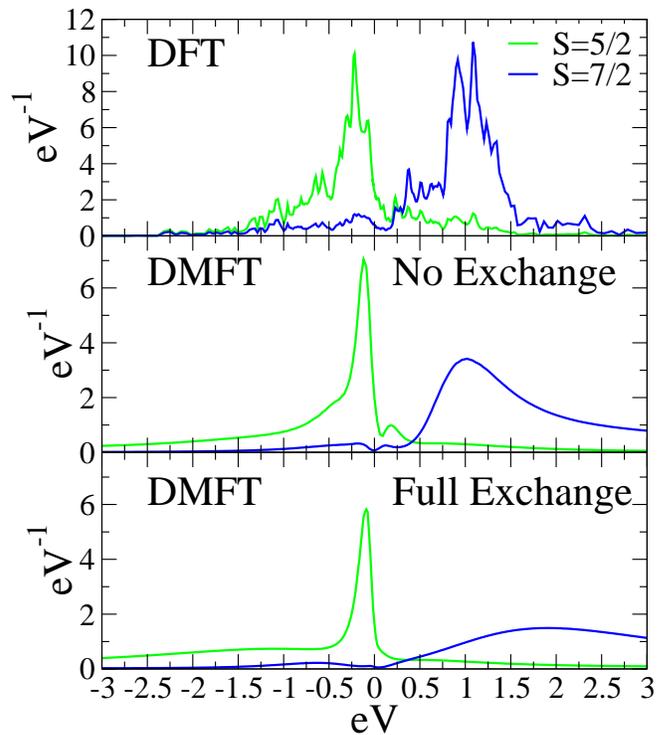}
\caption{ 
The spectral functions for the Pu $f$-electrons within DFT (top), DFT+DMFT
without exchange (middle), and DMFT+DFT with the full rotationally-invariant
exchange (bottom).
} \label{spec}
\end{figure}

We proceed by first exploring the qualitative effect of electronic correlations
on the local spectra, both with and without the rotationally-invariant exchange
interaction. The real-frequency spectral function is obtained by using the
maximum entropy method to analytically continue the imaginary time Green's
function measured in the CTQMC simulation.  The LDA spectrum displays a strong
spin-orbit splitting among the $S=\frac{5}{2}$ and $S=\frac{7}{2}$ states (see
figure \ref{spec}a).  The DFT+DMFT spectrum without exchange indicates
that spectral weight from low energies (ie.  near the Fermi energy) has
transferred to higher energies (see figure \ref{spec}b).  Including the full
exchange interaction reduces the spectral weight in the $S=\frac{5}{2}$
quasiparticle peak and hence the electronic coherence scale. Additionally, the
$S=\frac{7}{2}$  peak above the Fermi energy is broadened due to a multiplet
splitting. The spectrum with full exchange interaction is similar to the
three-peaked spectrum obtained by Shim \emph{et al.}\cite{Shim:2007B}.  The
$S=\frac{5}{2}$ has a central peak just below the Fermi energy and a peak near
$~-1eV$, while the $S=\frac{7}{2}$ states have a peak near $~-0.6eV$. These
features are much broader than those obtained by Shim \emph{et
al.}\cite{Shim:2007B}, but this is expected due to the limitations of the
maximum entropy method in obtaining the real-frequency data.

\begin{figure}[htb]
\includegraphics[width=\linewidth,clip= ]{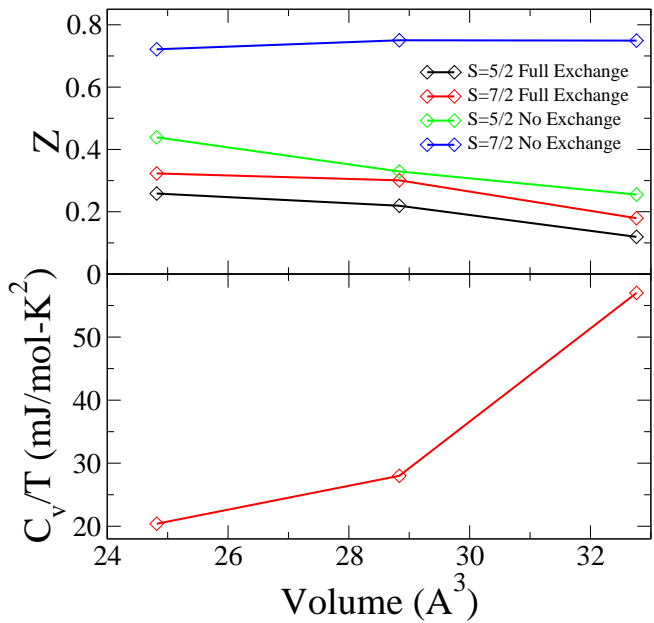}
\caption{ 
The quasiparticle weight $Z$ (top), and linear coefficient of the heat capacity
(bottom) as a function of volume. The heat capacity was calculated with the full
exchange interaction.
} \label{cv}
\end{figure}

The self-energy $\Sigma(i\omega)$ offers further insight into the nature of the
electronic correlations in $\delta$-Pu. The quasiparticle weight is determined
by the slope of the imaginary part of the self-energy (ie.
$Z={1}/{(1-\frac{\partial\Sigma(i\omega)}{\partial i\omega})}$), and within
DMFT the effective mass of the electrons due to electronic correlations is the
inverse of the quasiparticle weight (ie.  $\frac{m^*}{m}=\frac{1}{Z}$).  In the
absence of exchange, the $S=\frac{7}{2}$ states are very weakly correlated,
having an average $Z\approx0.7$ while the $S=\frac{5}{2}$ states have an
average $Z\approx 0.41$ for the equilibrium volume of $\delta$ Pu (see figure
\ref{cv}a). Weak correlations for the  $S=7/2$ states is expected given that in
the absence of Hund's coupling the j-j coupling scheme is adequate, and hence
the $S=\frac{7}{2}$ states are nearly empty.  Alternatively, the
$S=\frac{5}{2}$ states are much closer to a non-zero integer filling and are
moderately correlated.  When the exchange is included, the quasiparticle weight
is substantially decreased, resulting in $Z_{5/2}=0.26$, $Z_{7/2}=0.32$ for the
equilibrium volume.  The exchange interaction pushes Pu towards intermediate
coupling, where the $S=7/2$ states are more mixed into the ground state and
hence they becomes heavier.  With increasing volume, the kinetic energy
decreases and hence correlations increase resulting in a smaller $Z$. The only
exception is the $7/2$ orbital in the absence of exchange, where the occupation
slightly decreases with increasing volume.

Having established the effect of correlations on the spectra and the importance
of the full rotationally-invariant exchange interaction, we proceed to explore
the heat capacity and the magnetic susceptibility.  Within Fermi liquid
theory, the linear coefficient of the heat capacity is given by
$\gamma=\frac{2\pi k_B^2}{3} \sum_{\alpha}
\frac{\rho_{\alpha}(0)}{Z_{\alpha}}$, where $\alpha$ runs over all orbitals,
$\rho_\alpha$ is the local density of states, and $Z_\alpha$ is the
corresponding quasiparticle weight.  The heat capacity as a function of volume
is shown in Fig.~\ref{cv}.  As the volume is increased, the heat capacity
increases due to the fact that the quasiparticle renormalization amplitude $Z$
decreases and the spectral density at the Fermi energy increases.  The
predicted heat capacity for the equilibrium volume of $\delta$-Pu is $20.4
\frac{mJ}{mol-K^2}$.  The difference between the predicted value of the heat
capacity and the experimentally measured values of $35-55 \frac{mJ}{mol-K^2}$
may be be due to several factors.  Given that the $f$-electron spectral
function is extremely steep in the vicinity of the Fermi level (see Figure
\ref{spec}), it is clear that the value of $\gamma$ is sensitive
to small changes in the Fermi energy. Therefore, approximations in the
calculation, such as the local density approximation to the Kohn-Sham
potential, may have a non-negligible influence on the heat capacity.  The
second potential cause of this difference might be that the electron-phonon coupling
may further renormalize the hoppings and this is not included in our
calculation.  Given that our prediction is smaller than experiment, inclusion
of electron-phonon coupling would improve agreement with current experiments.
It should be noted that the largest volume point is not yet in the Fermi liquid
regime, as demonstrated below, and hence the Fermi liquid formula for the heat
capacity should only be considered as an estimate for the largest volume.

\begin{figure}[htb]
\includegraphics[width=\linewidth,clip= ]{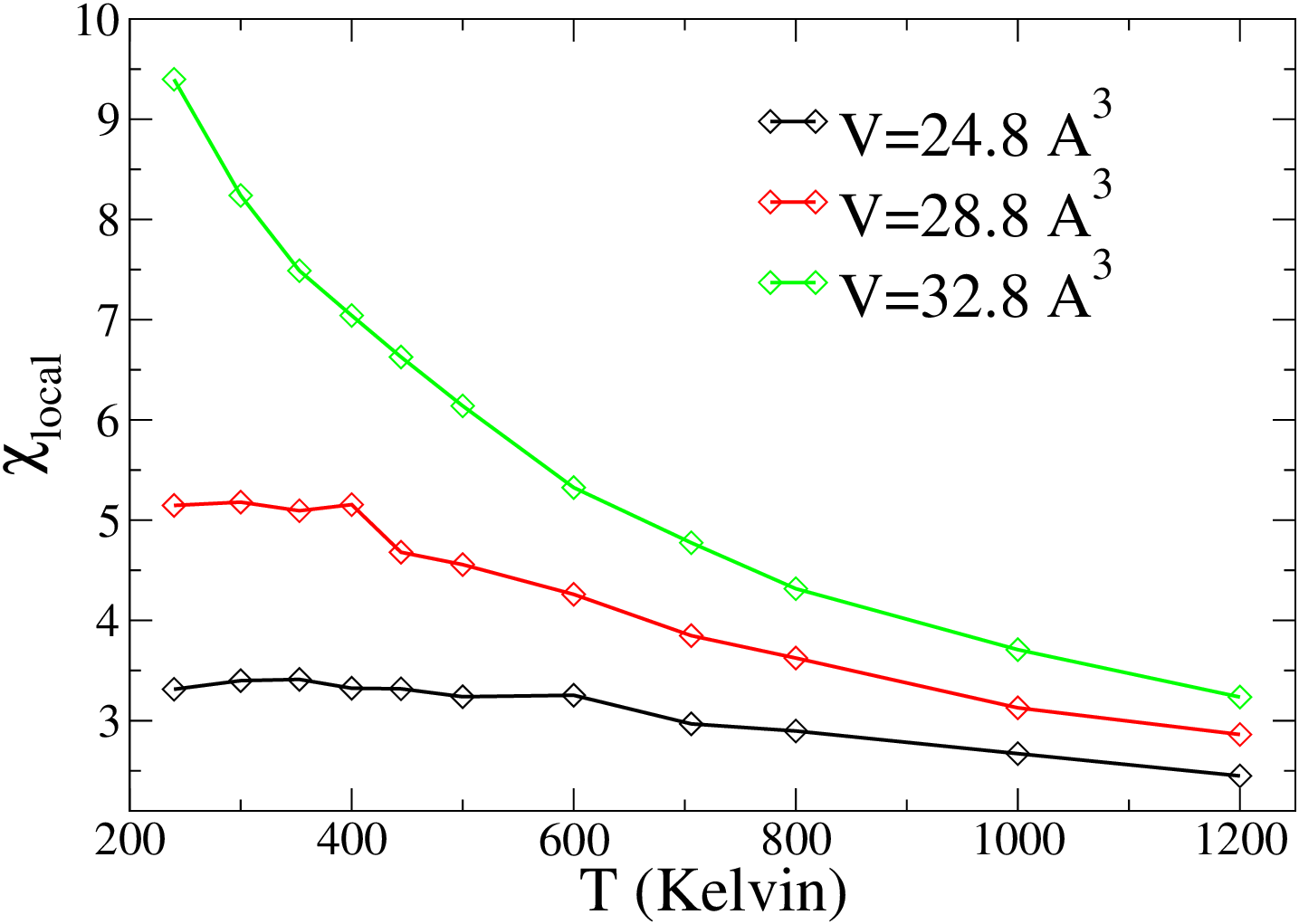}
\caption{ 
Local magnetic susceptibility as a function of temperature for different volumes.
} \label{chi}
\end{figure}

The local magnetic susceptibility is calculated as a function of temperature
for different volumes using the following expression: 

$\chi_{local}=\int^\beta_0{d\tau\langle M(\tau)M(0) \rangle}$

where $M=L+2S$, $L$ is the total orbital angular momentum, and $S$ is the total spin angular momentum.
For the equilibrium volume of Pu, the susceptibility is
relatively flat (ie. Pauli-like) below $600K$ and diminishes at higher
temperatures.  This behavior is consistent with experimental measurements. Our
calculations predict that the reason that magnetic moments are not seen in
$\delta$ Pu is because the system is coherent and the moments are therefore
screened. This predicted behavior is in agreement with experimental
observations\cite{Lashley:2005,Heffner:2006,Heffner:2006B,Heffner:2007,Curro:2004}.
As the volume is increased by $16\%$, the low-temperature Pauli-like
contribution has been renormalized to higher values and transitions to a
Curie-like behavior around $400K$.  The increase in the Pauli contribution is
consistent with the decrease in $Z$ as the volume is increased, and this
illustrates an enhancement of the electronic correlations.  As the volume is
further increased by an additional $16\%$, the susceptibility is Curie-like to
the lowest temperatures reached in this study.  These results clearly indicate
a decoherence of the electrons. As the volume is increased, the coherence
energy of the electrons decreases and therefore a transition from coherent to
incoherent behavior (ie.  Pauli-like to Curie-like) occurs at increasingly
smaller temperatures.  This signifies that a local magnetic moment has emerged
at progressively lower temperatures as the volume is increased.

The general behavior observed in these magnetic susceptibility calculations is
consistent with experiments. First, doping Pu with Americium causes the Pu lattice
to expand, effectively increasing the volume. The Pauli contribution of the Pu
atom is shown to increase as the Americium content is increased and the
effective volume is increased\cite{Mccall:2008}.  This is consistent with the enhancement of the
Pauli term that we observed.  Secondly, when hydrogen is doped into the system,
the \emph{fcc} Pu lattice expands by more than $50\%$ and  Curie-like behavior
is measured for the magnetic susceptibility down to temperatures of $50K$\cite{Aldred:1979}. This is
qualitatively consistent with the largest volume expansion in our calculations.

\begin{figure}[htb]
\includegraphics[width=\linewidth,clip= ]{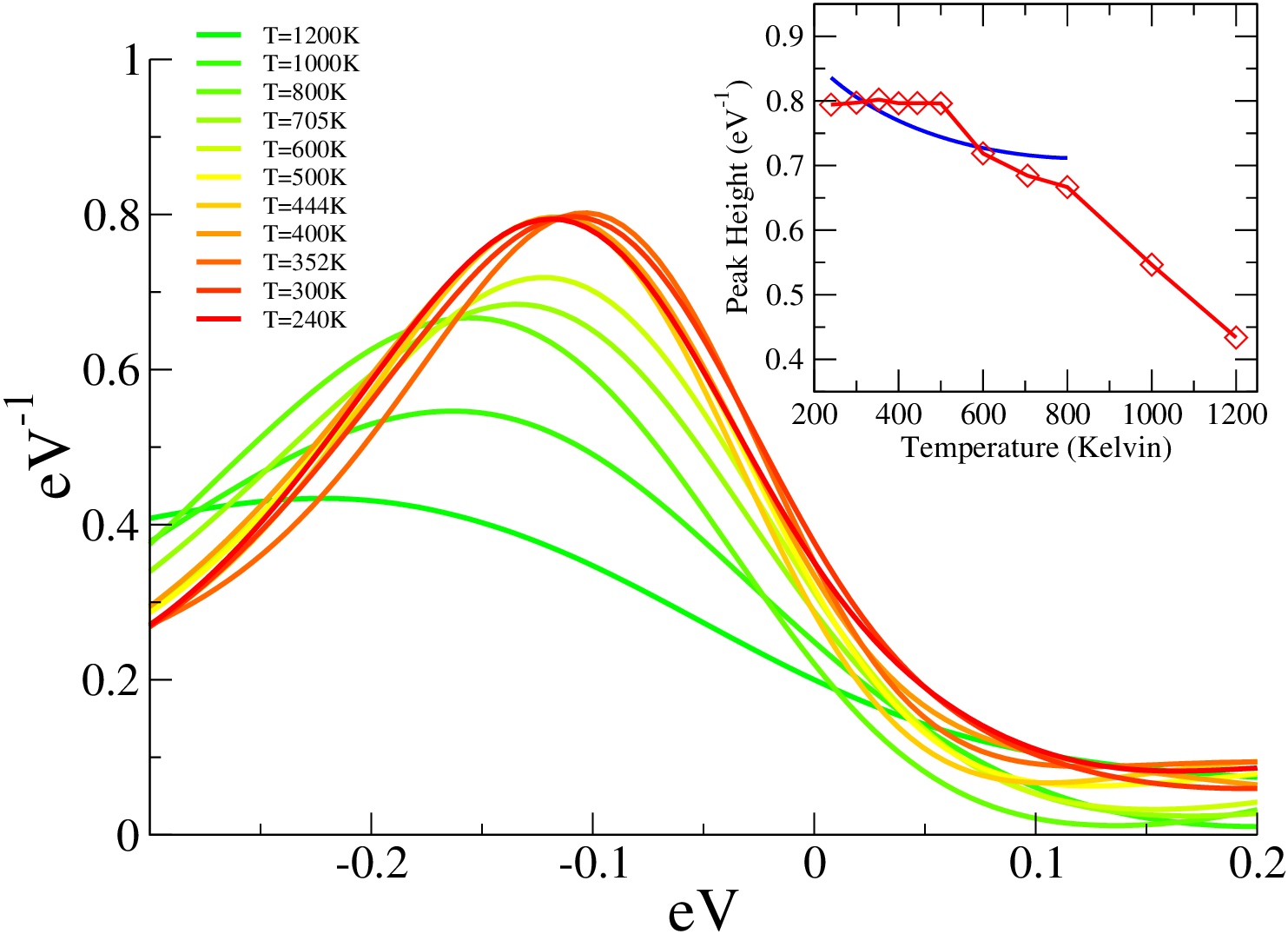}
\caption{ 
The temperature dependence of the $S=\frac{5}{2}$ $f$-electron spectral function using
the rotationally-invariant exchange interaction. The inset displays the peak height as a function of temperature.
The red points correspond to our data, while the blue curve is the parametrization from Ref. \onlinecite{Yang:2008}
assuming an onset temperature of $800K$.
} \label{tdspec}
\end{figure}

When the electronic behavior departs from the Fermi liquid theory and enters
the incoherent regime, there is a clear signature in the spectra.  In
Fig.~\ref{tdspec}, we show the temperature dependence of the local spectral
function at ambient pressure.  At high temperatures, the spectrum is diffuse.
As the temperature is decreased, a quasiparticle peak continually builds and
eventually saturates at $T=500K$. The coherence temperature may be defined as
the temperature at which the quasiparticle peak nears saturation, and we define
$75\%$ saturation to be the onset of coherence.  Hence our estimation for the
coherence temperature is approximately 800$\,$K, consistent with
Ref.\onlinecite{Shim:2007B}.

The inset of Fig.~\ref{tdspec} shows the temperature dependence of the height
of the quasiparticle peak. Notice that this behavior is very different from the
temperature dependence in heavy-Fermion compounds \cite{Shim:2007} recently
parametrized in Ref.~\onlinecite{Yang:2008}. For comparison, we plot the best
fit of the parametrization of Ref.~\onlinecite{Yang:2008} to our data. The
inability of this parametrization to describe our data is due to
the mixed-valence nature of Plutonium.

In conclusion, we have performed approximation-free DMFT calculations including
the rotationally-invariant exchange interaction for $\delta$-Pu. The efficient
CTQMC algorithm has allowed us to reach both high temperatures and
temperatures below ambient.  The quasiparticle weight for the $S=\frac{5}{2}$
states of $\delta$-Pu is found to be $Z=0.25$, indicating the presence of
appreciable electronic correlations.  Calculation of the magnetic
susceptibility indicates Pauli-like behavior for the equilibrium volume of
$\delta$-Pu, in support of experimental measurements. This indicates that in
$\delta$-Pu the moments are screened.  Expanding the volume causes the
electrons to crossover from coherent to incoherent behavior at increasingly
lower temperatures.  This crossover is illustrated in the temperature
dependence of the spectra. The prediction of incoherent electronic behavior, manifested as 
Curie-like behavior in the magnetic susceptibility, is consistent with experimental measurements
in PuH$_2$. The importance of the inclusion of the rotationally
invariant exchange is illustrated in the spectrum and in the reduction of the
quasiparticle renormalization amplitude from $Z=0.41$ to $Z=0.25$.

Acknowledgment: This work performed under the auspices of the U.S.
Department of Energy by Lawrence Livermore National Laboratory under
Contract DE-AC52-07NA27344.  Computer time was awarded by the Atlas
grand challenge program at LLNL.  We acknowledge useful conversations
with M. Manley, S. Mccall, and K. Moore. We thank K. Moore for revising 
the manuscript.

\bibliography{prl}
\end{document}